\documentclass[aps,prb,showpacs,twocolumn,floats,epsfig,pdflatex]{revtex4-1}
\usepackage{epsfig}
\usepackage{epstopdf}
\usepackage{amsmath}
\usepackage{amsfonts}
\usepackage{graphicx}
\usepackage{amssymb}
\usepackage{amsbsy}
\usepackage{subfigure}

\newcommand{\cs}{$\clubsuit$ \bf}

\begin{document}

\title {Quantum phases of hardcore bosons in two coupled chains: A DMRG study}

\author{Bradraj Pandey$^{(1)}$, S. Sinha$^{(2)}$, and Swapan K. Pati$^{(1,3)}$}
\affiliation{$^{(1)}$Theoretical Sciences Unit, Jawaharlal Nehru Centre for Advanced Scientific Research, Jakkur P.O., Bangalore-560064, India.
\\ $^{(2)}$ Indian Institute of Science
Education and Research-Kolkata - Mohanpur, Nadia 741252, India.\\ $^{(3)}$
New Chemistry Unit, Jawaharlal Nehru Centre for Advanced Scientific Research, Jakkur Campus, Bangalore 560064, India
 }

\date{\today}

\begin{abstract}
We consider hardcore bosons in two coupled chain of one dimensional lattices at half filling with repulsive intra-chain interaction and inter-chain attraction. This can be mapped on to a coupled chain of spin-1/2 $XXZ$ model with inter chain ferromagnetic coupling. We investigate various phases of hardcore bosons (and related spin model) at zero temperature by density matrix renormalization group method. Apart from the usual superfluid and density wave phases, pairing of inter chain bosons leads to the formation of novel phases like pair-superfluid and density wave of strongly bound pairs. We discuss the possible experimental realization of such correlated phases in the context of cold dipolar gas.  

\end{abstract}


\maketitle

\section{Introduction}
After successful experimental realization of dipolar Bose-Einstein-condensation (BEC)
of  $^{52}Cr$\cite{Thierry}, $^{164}Dy$\cite{Lu},
and Rydberg atoms\cite{weim}, possibility of finding exotic phases like superfluid,
pair-superfluid, supersolid, pair-supersolid, charge density wave and phases involving 
quantum magnetism\cite{BoYan} have increased tremendously. Usually, bosons can form superfluid by
condensation of bosonic particles to a single ground state, whereas fermionic superfluidity
 in superconductors and in cold atoms\cite{Markus,regal} occurs due to the formation of pairs.
For sufficiently strong attractive interactions, bosons can also form pairs 
which leads to the formation of `pair-superfluidity' of bosons\cite{Ni}. 
Pair-superfluidity can be realized in cold atom systems by interspecies
 attractive interactions\cite{kuk,chi}, bilayer dipolar systems\cite{Tre,safi,Amacia}, 
and through Feshbach resonance\cite{cat}.
Theoretically `pair-superfluidity' has also been studied in models with 
correlated hopping\cite{Hong}.

A supersolid phase is described by simultaneous 
existence of crystalline order and superfluid order in  the system. Various experimental 
and theoretical studies have been carried out for finding supersolidity
\cite{kim,Duk,dlkov,Melko,Sengupta,sinha,jiang,adam,N.Henke,F.Cinti,Xi}. Interestingly, 
pair-supersolid (PSS) is defined as a phase where one finds 
simultaneous existence of pair-superfluidity and modulation in density,
with vanishing single-particle superfluidity\cite{Tre,safi,w.zhang,chi}.
Bilayer dipolar systems  provide existence of pair-superfluid (PSF) and
pair-supersolid (PSS) phases\cite{Tre,safi}. The 
possibility of pair-supersolidity in bilayer dipolar gas with polarised dipoles has been also investigated\cite{Tre}, where the existence of PSF and PSS phases are shown by solving an effective Hamiltonian of pairs in the strong coupling limit.

 Trefzger, $et \ al.$, have looked at polarized dipolar
 particles in two decoupled 2D layers, in the  presence of repulsive interactions in the planes and 
attractive interactions between the two layers. They have shown the  existence of PSS and PSF
 phases by solving the effective extended Bose-Hubbard Hamiltonian in the low-energy
 subspace of pairs, by means of a mean-field Gutzwiller approach and exact
 diagonalization methods\cite{Tre}.
The PSF and PSS phases have  also been studied in a two-species Bose-Hubbard model
in a two-dimensional square lattice with on-site intraspecies
repulsions and interspecies attractions\cite{chi}.

 Low dimensional quantum systems are quite unique, as in reduced dimension,
 quantum fluctuations destroy the  true long range order (LRO). Instead, the 
low dimensional systems quite often show quasi long range order (QLRO). Incidentally,
for the system to show QLRO, the equal-time correlation functions, 
$\langle C^+(X)C(0)\rangle$,(where $X$ is the distance) would decay algebraically. However, if the correlation function decays exponentially, the system is believed to show short range order (SRO)\cite{giamarchi}. The transition between superfluid to Mott insulator
 in one dimension at commensurate density is a BKT type transition, and the transition point can be determined 
by a Luttinger liquid parameter, K\cite{T.D,Till,G.Roux,M.A,S.E,Meetu}. The Luttinger liquid parameter
 can be extracted from exponent of correlation functions. 
For bosonic low-dimensional systems, there have been studies where
a number of phases, namely, superfluid, supersolid, and pair-superfluid, phases have been 
reported\cite{lmat,lmazza,bat,swap,tapan1,tapan,mish,Anzi,arg,gs,mg,as,man}.
In low dimension, quite a few  interesting studies  in paring phenomena have been carried out.
Paired superfluid and counterflow superfluidity in one-dimension can exist in binary mixture of bosons with equal density\cite{Anzi}.
Studies on phases of the dipolar 
bosonic  gases in unconnected neighboring one-dimensional
systems have also been carried out\cite{arg}.
Parallel stack of  one-dimensional hard-core bosons in  optical lattices have been studied,
 by using bosonization and quantum Monte Carlo methods\cite{as},
 where superfluids, super-counterfluids (SCF), and checker-board (CB) phases
 from composite  particles from different tubes are shown. 
In a recent study \cite{man} of a two-leg 
ladder system with attractive onsite and repulsive interchain nearest-neighbor 
interactions, phases like atomic superfluid,
dimer superfluid and dimer rung insulator are found by imposing the onsite three-body constraint.

Motivated by recent experimental progress on dipolar gas, we consider hardcore bosons with dipolar interactions on two coupled one dimensional chains at half-filling. Dipoles are oriented in such a way that it generates a nearest neighbour intra-chain repulsion and onsite inter-chain attraction. In this system, inter-chain attraction can induce pairing between the bosons in two chains and intra-chain repulsion can break the translational symmetry which leads to the formation of density ordering. In this work, we mainly focus on the formation of various phases due to the interplay between these two ordering. The remaining part of the paper is organized as follows. In sec.II we describe the model and its connection to an equivalent spin model. Various phases of this bosonic ladder (and spin chain) with different ordering are discussed. The results obtained from DMRG calculations are presented in details in Sec.III. Different phases and their transitions are described in separate subsections. Finally, we summarise all our results in Sec.IV.

\section{The Model}
%********************************Fig.1.*****************************
\begin{figure}[h]
\rotatebox{0}{\includegraphics*[width=\linewidth]{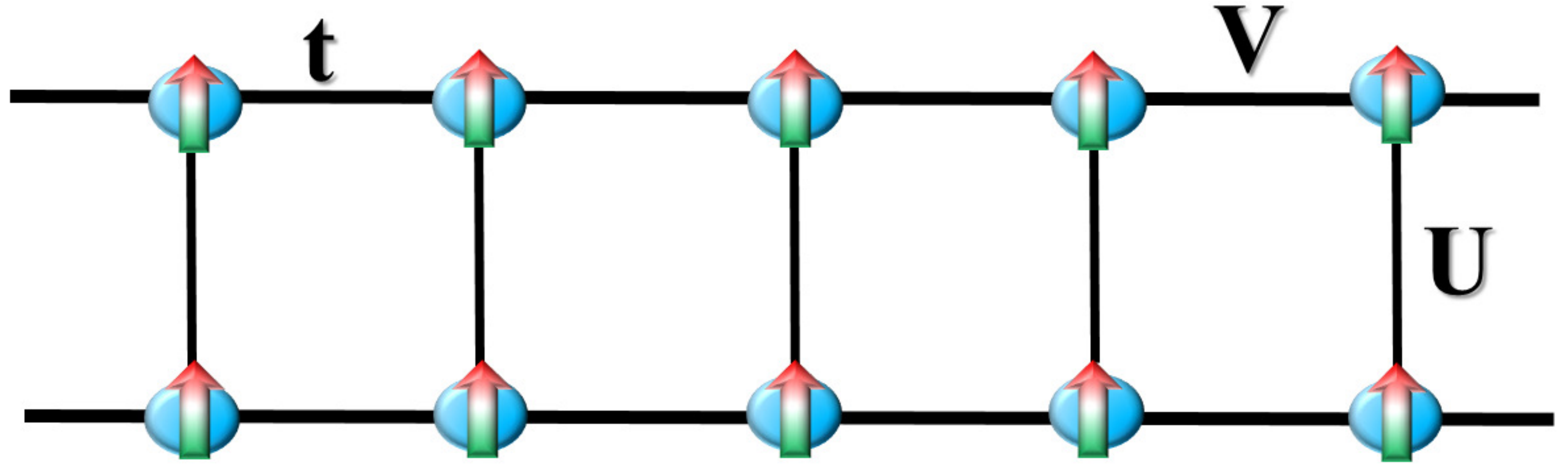}}
\caption{Schematic of the two chains with dipolar bosons. There is nearest-neighbour repulsive interaction $V$, and 
nearest-neighbour hopping parameter $t$, in each of the  chain. Both chains are coupled with onsite attractive interaction $U$, while there is no  hopping between the two chains.}
\label{fig1}
\end{figure}
%*******************************************************************

We consider hardcore bosons in two coupled chain of one dimensional lattices at half filling 
 with dipolar interaction as depicted in Fig.1. The anisotropic part of  dipolar interaction is proportional to $\left(1-3cos^2(\theta) \right)$, 
where $\theta$ is the  angle between the dipoles. We consider that the dipoles are polarized perpendicular to the chains (as shown in Fig.1). Thus,
the dipolar interaction is repulsive when dipoles are in the same chain, while, the dipoles of different chains which are at the same lattice site attract each other. 
The effective Hamiltonian of the  system, without taking into account the  inter chain hopping, can be written as,
\begin{eqnarray} 
H= & & -t \sum_{\alpha,<i,j>} \left(b^{\dagger}_{\alpha,i} b_{\alpha,j} + h.c\right) + V \sum_{\alpha,<i,j>}
 \hat{n}_{\alpha,i} \hat{n}_{\alpha,j}\nonumber  \\
& -& U\sum_i \hat{n}_{1,i} \hat{n}_{2,i}
\end{eqnarray}
\noindent where $\alpha=1,2$, is the chain index, $t$ is the hopping
 term within the chains, $V$ is the strength of intra-chain nearest-neighbour repulsion and $U$ is the strength of inter-chain onsite attraction.
For simplicity, we truncate the long range dipolar interaction and consider only nearest-neighbour intra-chain repulsion and onsite inter-chain attraction. The physical states of a hardcore boson are restricted by the condition $b_i^{\dagger 2}|0\rangle =0$.
The number states of a hard core boson is equivalent to $s^z$ states of a spin-1/2 particle by the mapping
$ (|1\rangle \rightarrow |\uparrow\rangle$ and $|0\rangle \rightarrow |\downarrow\rangle )$.
The creation, annihilation operators of a hard core boson can be represented by the spin-1/2 operators as follows,
$s^+_i \to b^{\dagger}_i$, $s^-_i \to b_i$ and $s^z_i \to  \ n_i \ - \ 1/2$.  
The final spin Hamiltonian turns out to be a coupled chain of spin-1/2 $XXZ$
model with inter chain ferromagnetic coupling,
\begin{eqnarray}
H= & & -t \sum_{\alpha,<i,j>} \left(s^{+}_{\alpha,i} s^{-}_{\alpha,j} + h.c\right) + V \sum_{\alpha,<i,j>} 
s^{z}_{\alpha,i} s^{z}_{\alpha,j}  \nonumber \\ 
& - & U\sum_i s^{z}_{1,i} s^{z}_{2,i}. 
\end{eqnarray}

In the spin Hamiltonian (eq.2), which is obtained from bosonic Hamiltonian (eq.1), we impose the constraint $\sum_{i} s^{z}_{\alpha, i} = 0$. 

%**************************Fig.2.********************************** 
\begin{figure}[h]
\rotatebox{0}{\includegraphics*[width=\linewidth]{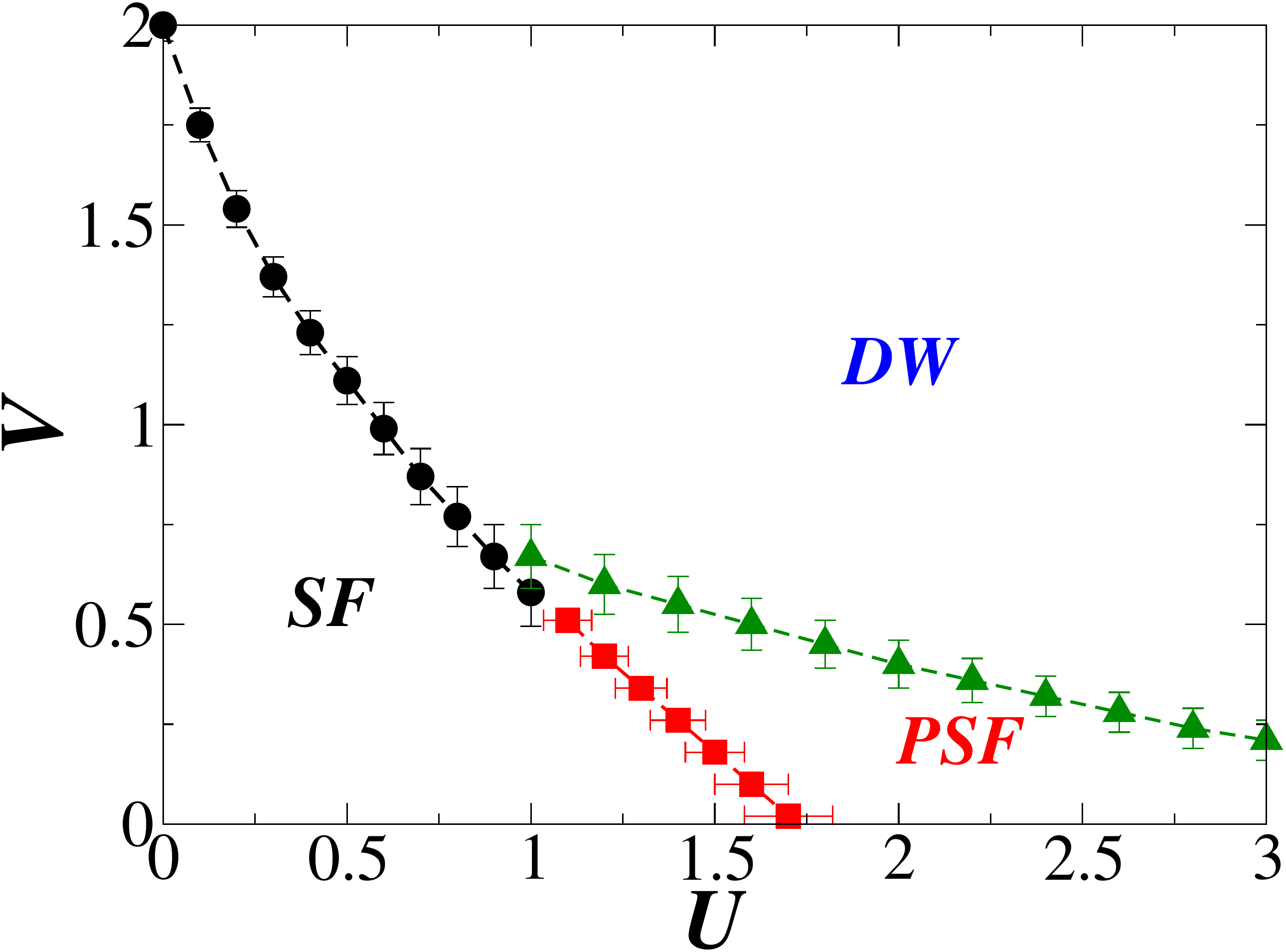}}
\caption{Two dimensional phase diagram in the phase space of two parameters, $U$ and $V$. 
The phase diagram is quite rich with  phases, namely, Superfluid (SF),
 Pair-superfluid (PSF) and Density wave (DW) phases. }
\label{fig7}
\end{figure}
%*********************************************************************

In this model, we scale all the energies by the hopping strength, $t$ and set $t=1$ to obtain the complete phase diagram in the $U-V$ plane. 
For $U=0$, the above model becomes equivalent to two decoupled $XXZ$ spin $1/2$
 chains which can be solved exactly and studied extensively\cite{giamarchi}.
 This model undergoes a quantum phase transition
 to antiferromagnetic phase above the critical coupling value, $V=2$.
 Similarly, hardcore bosons with nearest neighbour repulsion exhibits a transition from superfluid to density wave. 
Superfluid and density wave phases can be characterized by following correlation functions,
\begin{eqnarray}
C_{\alpha}(r) & = & \langle b^{\dagger}_{\alpha, i} b_{\alpha, i+r}\rangle,\\
G_{\alpha}(r) & = & \langle n_{\alpha,i} n_{\alpha,i+r}\rangle.
\label{corr_sfdw}
\end{eqnarray}   
For spin chain, corresponding correlations functions transform to $C_{\alpha}(r) = \langle s^{+}_{\alpha,i}s^{-}_{\alpha,i+r} \rangle$ and $G_{\alpha}(r) = \langle s^{z}_{\alpha,i} s^{z}_{\alpha,i+r} \rangle$.
In SF phase of bosons the correlation function $C_{\alpha}(r)$ shows power law decay
$\sim 1/r^{\alpha_s}$, where the exponent $\alpha_s$ can be determined from the 
Luttinger parameter\cite{giamarchi}.

We have calculated relevant quantities by varying the values of the parameters $U$, $V$
 and the phase diagram is shown in Fig.2.
For low values of $U$ and $V$, bosons in the two chains are almost decoupled and form a superfluid in each of the chains.
 In terms of spins, there will be quasi-long range order in the X-Y plane\cite{jiang}. In this case, 
the  effect of fluctuation is  quite large and there is no order along the z-axis.
For sufficiently large nearest neighbour interaction,
 density ordering develops in each chain which can be characterised
by the density-density correlation function $(-1)^{r}G_{\alpha}(r)$.
In DW phase, superfluidity vanishes and $C_{\alpha}(r)$ decays exponentially due to the appearance of an energy gap.
Attractive interaction between two chains induces pairing of bosons
 which can be analyzed from the correlation function of the pairs,
\begin{equation}
P(r) = \langle b^{\dagger}_{1,i}b^{\dagger}_{2,i}b_{2,i+r}b_{1,i+r} \rangle -\langle b^{\dagger}_{1,i}b_{1,i+r}\rangle
\langle b^{\dagger}_{2,i}b_{2,i+r}\rangle.
\label{corr_psf}
\end{equation}
For sufficiently large attractive interaction, $U$, and small repulsive interaction, $V$, a quasi
 `pair-superfluid'(PSF) state of bound pairs is formed. In this phase the correlation function, $P(r)$ shows QLRO but single particle superfluidity vanishes.
In the  large $U$ and  $V$ limit, the system forms strongly bound pairs of hardcore bosons
with density ordering of the pairs due to the strong nearest neighbour repulsion. 
This insulating density wave phase of pairs can be described by the wavefunction,
\begin{equation}
|PDW\rangle = \prod_{i} |0,0\rangle_{i}\prod_{j}|1,1\rangle_{j}
\label{wvfdw}
\end{equation}
where $i,j$ represent sites of two sublattices and $|n_{1},n_{2}\rangle_{i}$ is the number state of coupled chains at site $i$.   
 In terms of spin language, spins are ordered antiferromagnetically  in each of the chains, while spins  align ferromagnetically along the rung of the ladder.
This phase is similar to the `pseudo-gap' phase of superconductors, where phase coherence between the strongly bound pairs is absent.

\section{Results and Discussion}
To solve the above spin-Hamiltonian and to find various possible quantum phases in the
parameter space, we have used density-matrix renormalization group (DMRG)\cite{White,Schollwock} method.
We consider spin-1/2 at every site, varying the DMRG cut-off (max = m) from $250$ to $400$,
 for consistent results. Unless other wise stated, most of the results below 
are obtained with $m = 250$. We have used an open boundary condition for both the chains. 
We have  compared our DMRG results, namely  energy gap and  energy eigenvalues with results
from exact diagonalization, up to 28 lattice sites. 
We find the energies are comparable  up to five decimal places.
 To characterize different phases, we  have calculated spin-density, two points and 
four points correlation functions, and the corresponding structure factors.
For showing plots of correlation functions and structure factor, unless stated explicitly,
  we have considered each chain to be of length  $L/2= 160$,
 which amounts to the  total system size  $L=320$.
To determine an accurate phase boundary between different phases and to
minimize the finite size effect, we have done finite size scaling
of correlation lengths, structure factors and exponents of the correlation 
functions of the system with size (L) up to 384.

\subsection{SF to DW transition}
%********************************Fig.3****************************
\begin{figure}[h]
\rotatebox{0}{\includegraphics*[width=\linewidth]{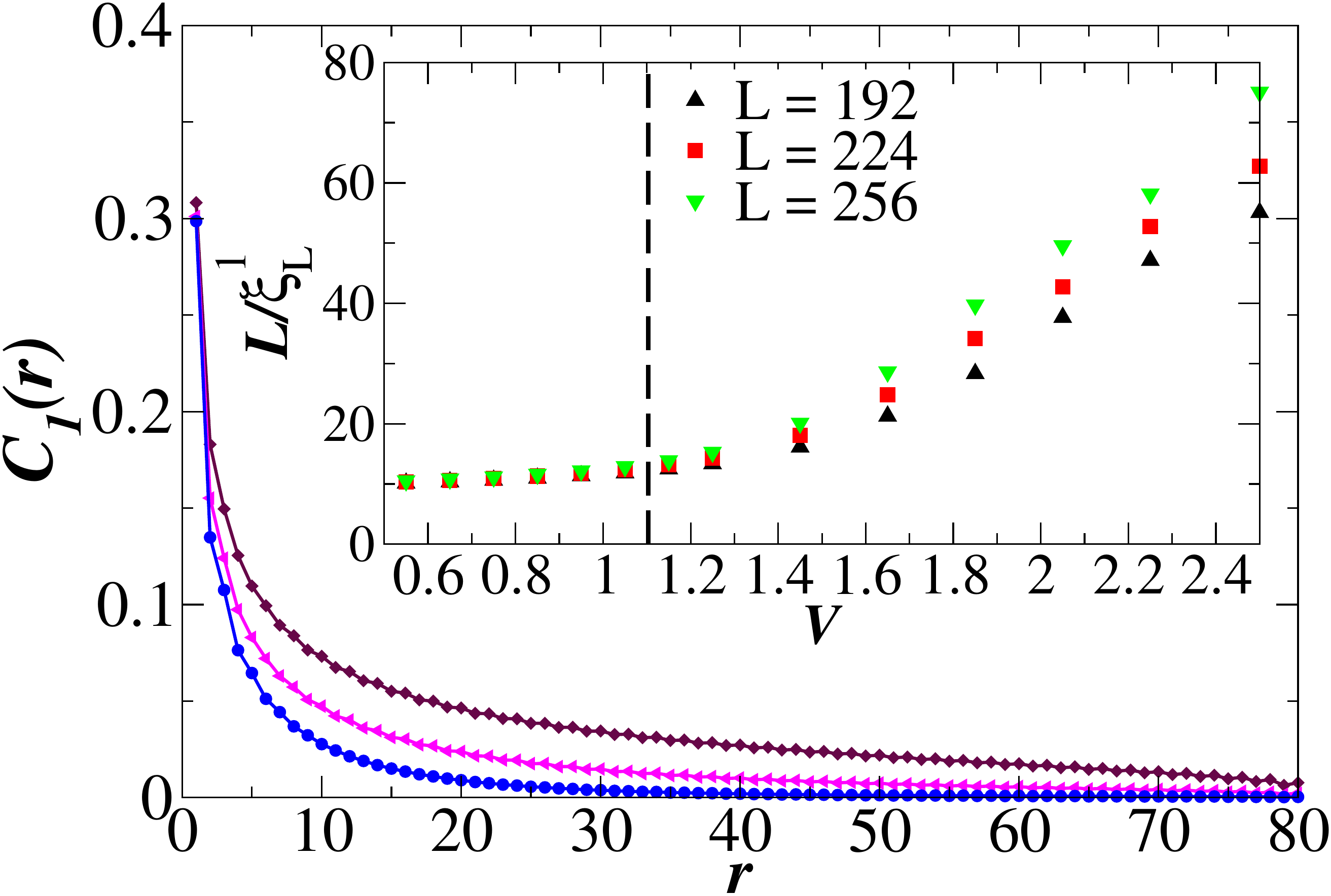}}
\caption{Plot of correlation function $C_1(r)$, as a function of $r$, at $U = 0.5$
 and different values of $V$ ($V = 0.4$ (square), $V = 1.0$ (triangle) and $V = 1.4$ (circle)).
 Inset shows scaling of  L/ $\xi_L^1$  as a function of $V$ for  $U = 0.5$.
Coalescence of the data points of different L  shows SF-DW transition at $V=1.1\pm0.05$.}
\label{fig3}
\end{figure}
%*********************************************************************

The quasi-superfluid order in terms of spin language  can be described as order
 in the XY-plane \cite{jiang}.
To calculate order along the XY-plane, we have calculated transverse spin-spin correlation function 
$C_{\alpha}(r)\  = \  \langle S^+_{\alpha,0}S^-_{\alpha,r}\rangle$,
 where $r$ is the distance from the middle of the chain. In Fig.3, 
we have shown the plot of the correlation function, $C_1(r)$ at $U=0.5$ and different values of $V$.
It shows, correlation function, $C_1(r)$,
 decays algebraically for  $V = 0.4$ and $1.0$, while, it has short range order for $V = 1.4$. 
The structure factor $C_1(k)= \frac{1}{(L/4)} \sum \exp(ikr)C_1(r)$ 
gives peak at $k = 0$ in the superfluid phase. For characterizing order along the z-axis (density wave),
 we have calculated the  correlation function $G_{\alpha}(r) =  \langle S^z_{\alpha,0}S^z_{\alpha,r}\rangle$.
In Fig.4, we have shown the plot of correlation function, $(-1)^rG_1(r)$, at $U=0.5$ and  different
values of $V$. The system has order along the z-axis  for  $V = 1.2$ and $1.4$, 
while it has short range order for $V = 0.4$. Due to the open-boundary condition in DMRG,
there exists some fluctuations in $G_1(r)$ close to the boundary.
 The structure factor $G_1(k)=\frac{1}{(L/4)} \sum \exp(ikr)G_1(r)$ gives peak at $k = \pi$ in the  density wave phase.  

%********************************Fig.4****************************
\begin{figure}[h]
\rotatebox{0}{\includegraphics*[width=\linewidth]{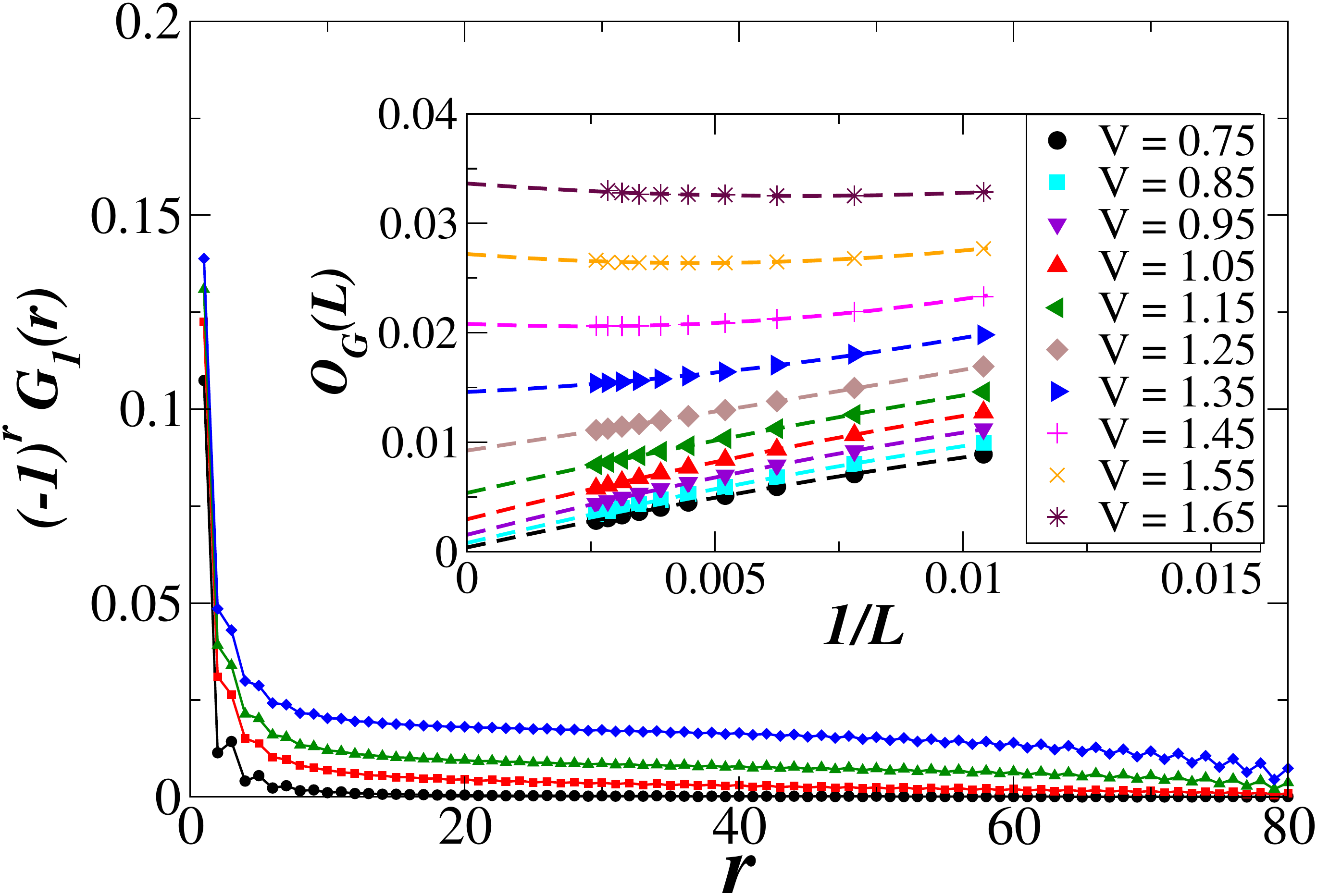}}
\caption{Plot of correlation function $(-1)^rG_1(r)$, as a function of $r$, at $U = 0.5$ 
 and different values of $V$ ($V = 0.4$ (circle), $V = 1.0$ (square), $V = 1.2$ (triangle) and $V = 1.4$ (diamond)).
 Inset shows, finite size scaling of  $O_G(L)$, at $U=0.5$, and for different values of $V$.}
\end{figure}
%*********************************************************************

 The transition between superfluid to gapped density wave  in one dimension 
is a BKT type transition. Thus the system opens up a gap very slowly, as it makes the transition 
 from SF to DW\cite{T.D,Till,G.Roux}.
 As energy gap and correlation length are related to each other ($G_L\sim 1/\xi_L$),
superfluid to density wave  transition can be shown by finite size scaling of the 
correlation length. The correlation length, is defined as\cite{Rpai,Rpai1,Meetu,G.Roux}
\begin{equation}
 \xi_L^{\alpha} = \sqrt{\frac{\sum_r r^2 C_{\alpha}(r) }{ \sum_r  C_{\alpha}(r)}}
\end{equation}
where $C_{\alpha}(r)\  = \  \langle S^+_{\alpha,0}S^-_{\alpha,r}\rangle$,
 is obtained by using the wave function of the system of length L. 
In the inset of Fig.3, we have plotted length dividing correlation length  $L/\xi_L^1$ versus
$V$, for $U = 0.5$. The  coalescence of data occurs at $V = 1.1\pm 0.05$ for different system sizes
 ($L = 192,224,256$). This indicates a transition from SF to DW at $V = 1.1\pm 0.05$.
  
Density wave order can  also  be characterized by a nonzero  static structure 
factor,  $O_G(L)=G_1(k=\pi)=\frac{1}{(L/4)}\sum_r (-1)^r G(r)$ \cite{bat,G.GBAT1,safi,D.ross}.
To obtain the thermodynamic value of $O_G(L)$, we have done finite size scaling
 for systems with length L up to $384$, by fitting the finite size $O_G(L)$ \cite{bat}
values with a function, $O_G$ + $O_1/L$  + $O_2/L^2$.
In the inset of Fig.4, we have plotted $O_G(L)$ as a function of 1/L  at $U=0.5$ and  different values of $V$.
From inset of the Fig.4, its appears that, the extrapolated value of $O_G(L)$, is finite for $V \gtrsim 1.05$.
On the contrary, for lower values of $V$, $O_G(L)$ decreases faster 
to very small values with increase in system size. This should have gone to zero in the thermodynamic limit, 
however, due to the BKT nature of the transition, the extraploated value of $O_G(L)$ goes to small nonzero values,
particularly near the critical region of SF-DW transition. In fact, due to the slow nature of the transition, 
from extrapolation of $O_G(L)$, it is difficult to exactly locate the phase boundary of the SF-DW transition. 
However, the significance of $O_G(L)$ plot is that it shows how the DW wave appears in the system, while 
going from a SF to DW phase. Note that, in the density wave phase, $G_1(r)$ decays exponentially to a 
non-zero value (except for some fluctuations near the boundary). Therefore, as shown in Fig.4, from correlation 
function, $(-1)^r G_1(r)$ and finite size scaling of $O_G$, we have estimated the density wave order 
in the system for $V = 1.1\pm0.08$, at $U=0.5$.

%********************************Fig.5****************************
\begin{figure}[h]
\rotatebox{0}{\includegraphics*[width=\linewidth]{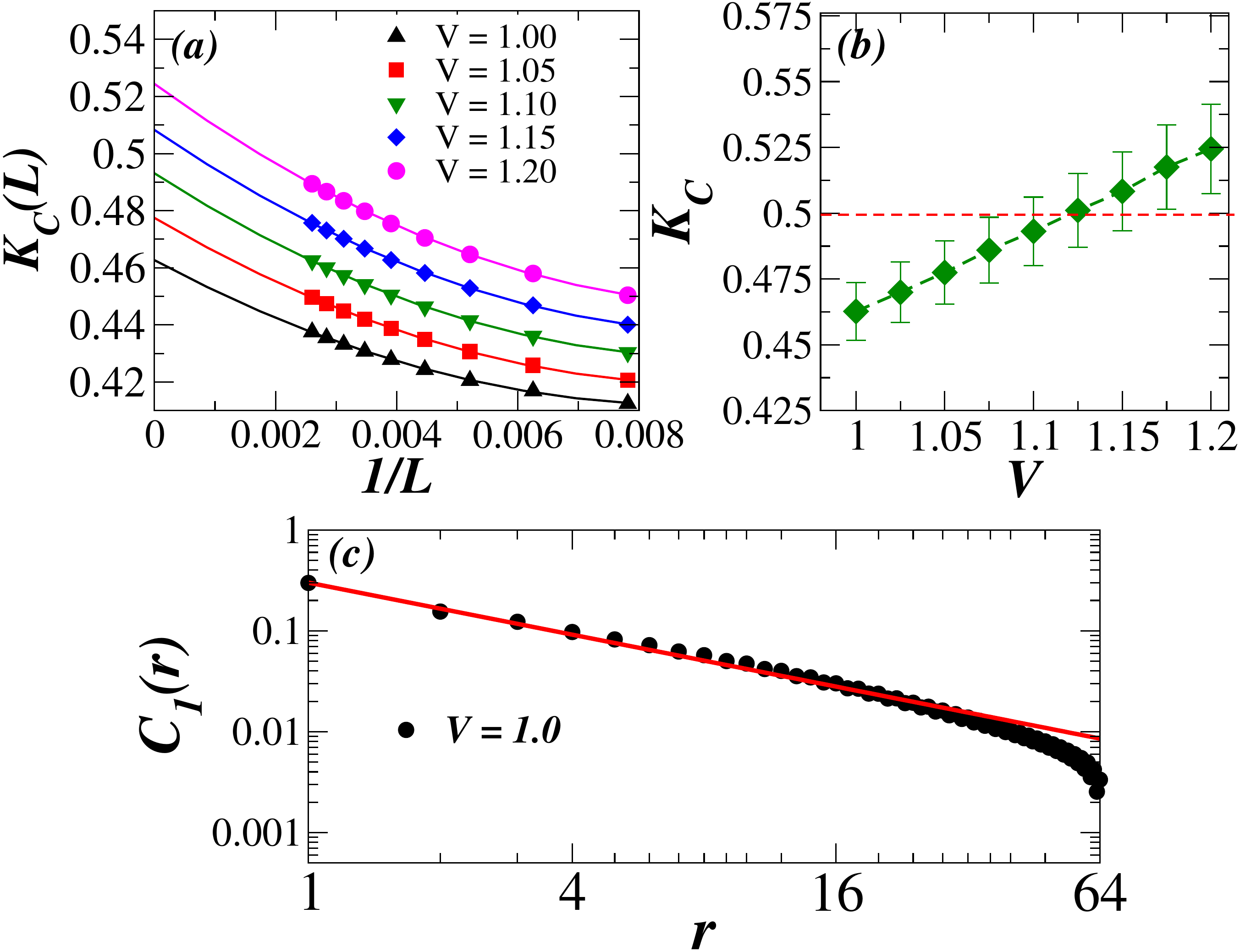}}
\caption{(a)Finite-size scaling of $K_C(L)$, at $U = 0.5$ and different values of $V$. 
(b) Plot of the extrapolated values of $K_C(L)$ vs $V$ for $U=0.5$,
showing SF to DW transition at $V=1.12\pm0.04$.
(c) Power law fitting of $C_1(r)$ for $V=1.0$, on a log-log scale.}
\label{fig4}
\end{figure}
%*********************************************************************

 As mentioned above, transition between SF to DW in one dimension is BKT type.
  The transition point can  also be determined by examining the critical
 exponent of the correlation function\cite{T.D,Till,Meetu}. Critical exponent 
 can be obtained by fitting the  correlation function with algebraic decay of 
 $C_1(r)=A/r^{2K_C}$ (as shown in Fig.5(c)).At the transition point ( from SF to DW), exponent ($K_C$) of the
 function $C_1(r)$ takes the value  $1/2$.
 The thermodynamic limit of $K_C(L)$ is obtained by extrapolating $K_C(L)= K_C+K_1/L+K_2/L^2$,
 where $K_1$ and $K_2$ are constants.
In Fig.5, we have shown SF to DW transition from $K_C$
 of the correlation function, $C_1(r)$ at $U=0.5$ and by varying $V$.  
In  Fig.5(a), we have shown extrapolation of $K_C(L)$, obtained from a power law fit of $C_1(r)$ for  
 different system sizes. 
 Extrapolation of $K_C(L)$ goes to  1/2   at $V = 1.12$ (inset of Fig.5(b)). This 
indicates a phase transition from SF to DW at $V = 1.12\pm0.04$ for $U=0.5$.
The error of $\pm0.04$ is the error in fitting of  $C_1(r)$ to the algebraic function.
In Fig.5(c), we have shown fitting of a correlation function, with $C_1(r)=A/r^{2K_C}$
for $V=1.0$ and with chain length $l=128$.
Due to the open boundary condition, fitting is not good near  the end of the chain.
Also  while going from a SF phase to a DW phase,  fitting error increases.
For $U=0$ and $V=2.0$, which is the transition point from SF to DW, we find $K_C=1/2\pm0.01$.
 while with increase in  $U$ near the SF-DW boundary, the error in fitting of $C_1(r)$ also increases slowly,
The transition points obtained from scaling of the  $L/\xi_L^1$ and  exponent $K_C(L)$
 are consistent with each other within the error bars indicated in the phase diagram.

\subsection{SF to PSF transition}
%********************************Fig.6****************************
\begin{figure}[h]
\rotatebox{0}{\includegraphics*[width=\linewidth]{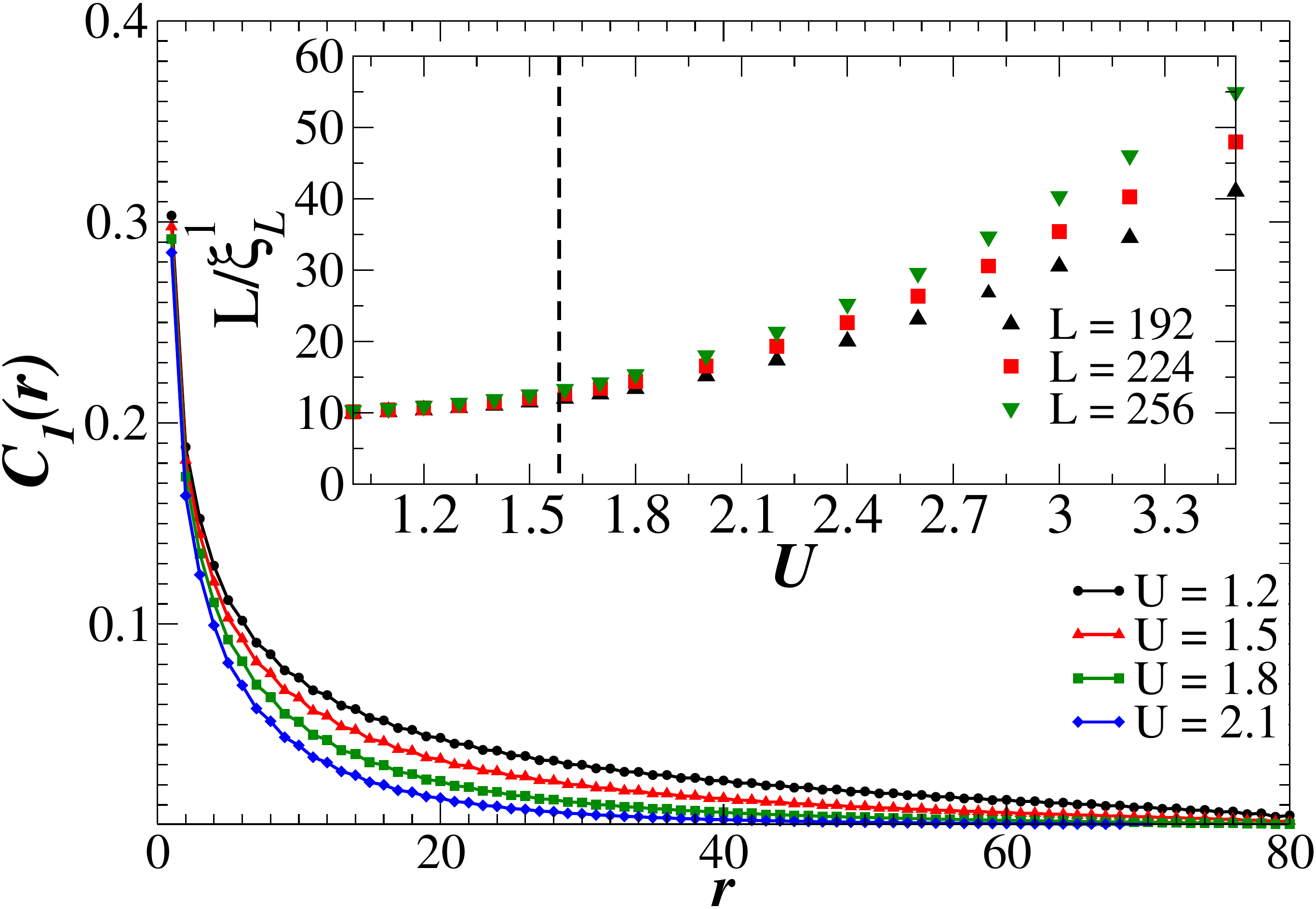}}
\caption{ Plot of correlation function $C_1(r)$, as a function of $r$, at $V = 0.1$ and different values of $U$.
In the inset, scaling of  L/ $\xi_L^1$  as a function of $U$ for  $V = 0.1$.
Coalescence of the data points of different system sizes shows SF-PSF transition at $U=1.6\pm0.1$.}
\end{figure}
%************************************************************

 With increase in  attractive interaction $U$ along the rungs of the ladder, hardcore bosons
  start making pairs along these  rungs. As a result, single particle superfluidity 
starts decreasing in each of the chains. 
For smaller values of repulsive interaction $V$, and sufficiently 
 large values of $U$, the system shows  BKT type transition from single particle
 superfluid phase to pair-superfluid phase\cite{Anzi,J.M.K}.
 In the PSF phase, single-particle spectrum opens up a gap. As a result the correlation function, $C_1(r)$,
 decays exponentially in this phase.
 As discussed in the case of SF to DW transition, here also, we estimate
  the SF to PSF transition from finite size scaling of correlation length $\xi^{\alpha}_L$. 
In Fig.6, we have plotted $C_1(r)$ vs $r$ at $V=0.1$ and different values of $U$.
 This plot, shows the transition from algebraic
 to exponential decay of, $C_1(r)$,
 as the system undergoes transition from  SF phase to PSF phase. In
the inset of Fig.6, we have plotted $L/\xi^1_L$ versus $U$.
The coalescence of data occurs at $U = 1.6\pm 0.1$ for different system sizes 
 ($L = 192,224,256$). This indicates transition from SF to PSF phase at $U=1.6\pm0.1$.
We find, generically, SF to PSF transition to be the slowest transition in the phase diagram.
The corresponding errors in finding the transition points have been indicated in the phase diagram.

%********************************Fig.7.*****************************
\begin{figure}[h]
\rotatebox{0}{\includegraphics*[width=\linewidth]{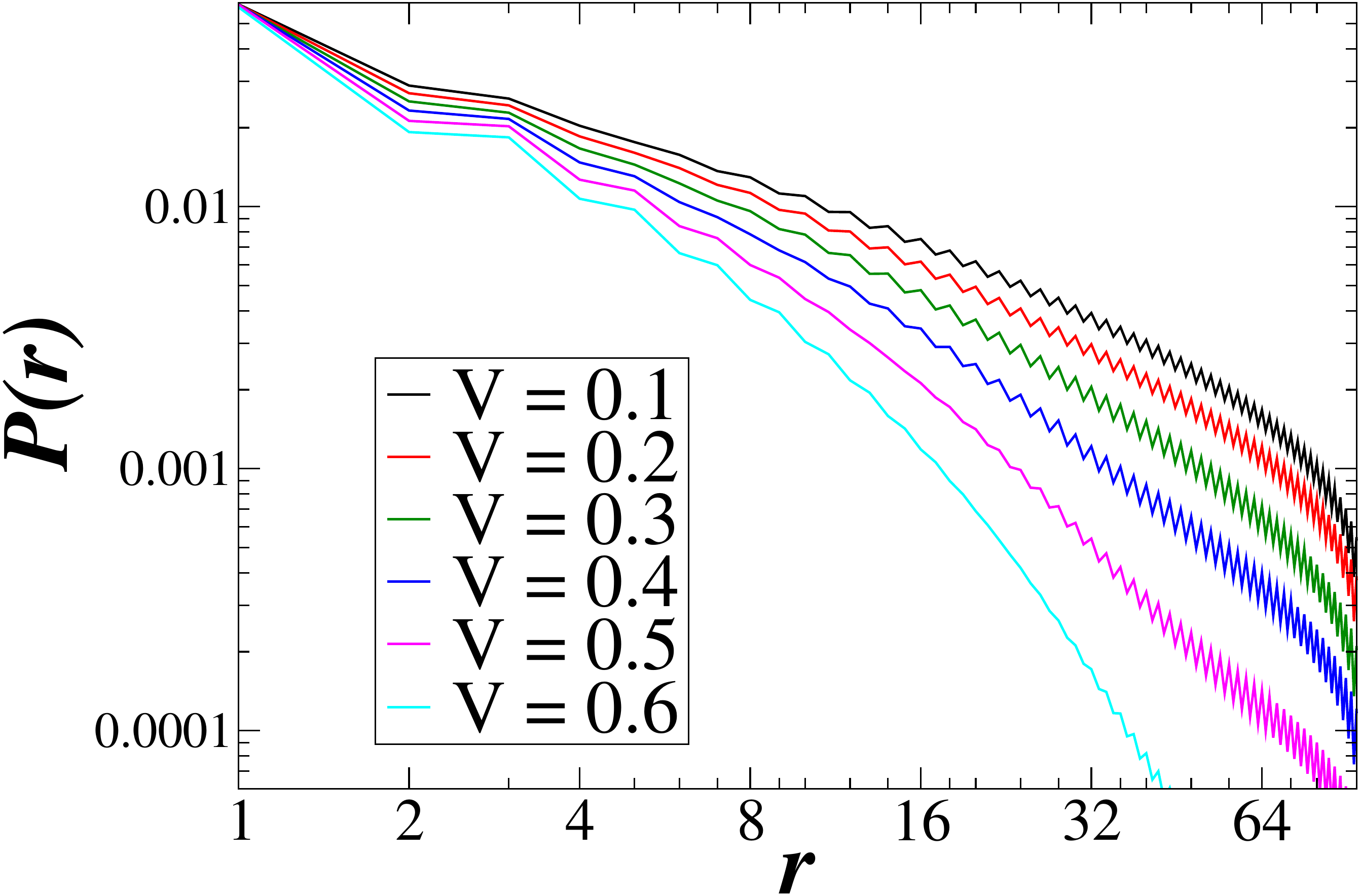}}
\caption{ Plot of pair-correlation functions $P(r)$, as a function of $r$,
 on a log-log scale, at $U = 2.0$ and  different values of $V$. The plot
shows PSF to DW transition  at $V=0.4\pm0.05$} 
\label{fig6}
\end{figure}
%********************************************************************* 

%*********************************************************************
\subsection{PSF to DW transition}

%********************************************************************* 
To characterize pair-superfluidity, we have  calculated the pair-correlation function, defined as,
$P (r)\  = \ \left \langle S^+_{1,0}S^+_{2,0} S^-_{1,r}S^-_{2,r}\right\rangle
- \left \langle  S^+_{1,0} S^-_{1,r}\right\rangle \left \langle S^+_{2,0} S^-_{2,r}\right \rangle $,
where $1$ and $2$ stand for chain indices of the  ladder and $r$
 is the distance from the middle site of the ladder.
We find pair-superfluidity in the system for  lower values of repulsive interaction $V$ and large enough values of attractive interaction $U$.
With increase in $V$, we find density wave in each of the  chains.
We also find that, in the  presence of large enough  $U$, 
density wave in each of the chains gets stabilized at much lower values of $V$,
 and become strongly correlated\cite{safi}. 
In the PSF phase, correlation function, $P(r)$, 
decays algebraically, while, in the density wave phase, it decays exponentially.
To reduce the finite size effect, we have calculated, $P(r)$ 
 by taking the total system size $L=384$ and with max  value $m=400$.
 In Fig.7, we have plotted the pair correlation function, $P(r)$, with $r$ in log-log scale, 
at $U=2$, and different values of $V$. 
 %********************************Fig.8.*****************************
\begin{figure}[h]
\rotatebox{0}{\includegraphics*[width=\linewidth]{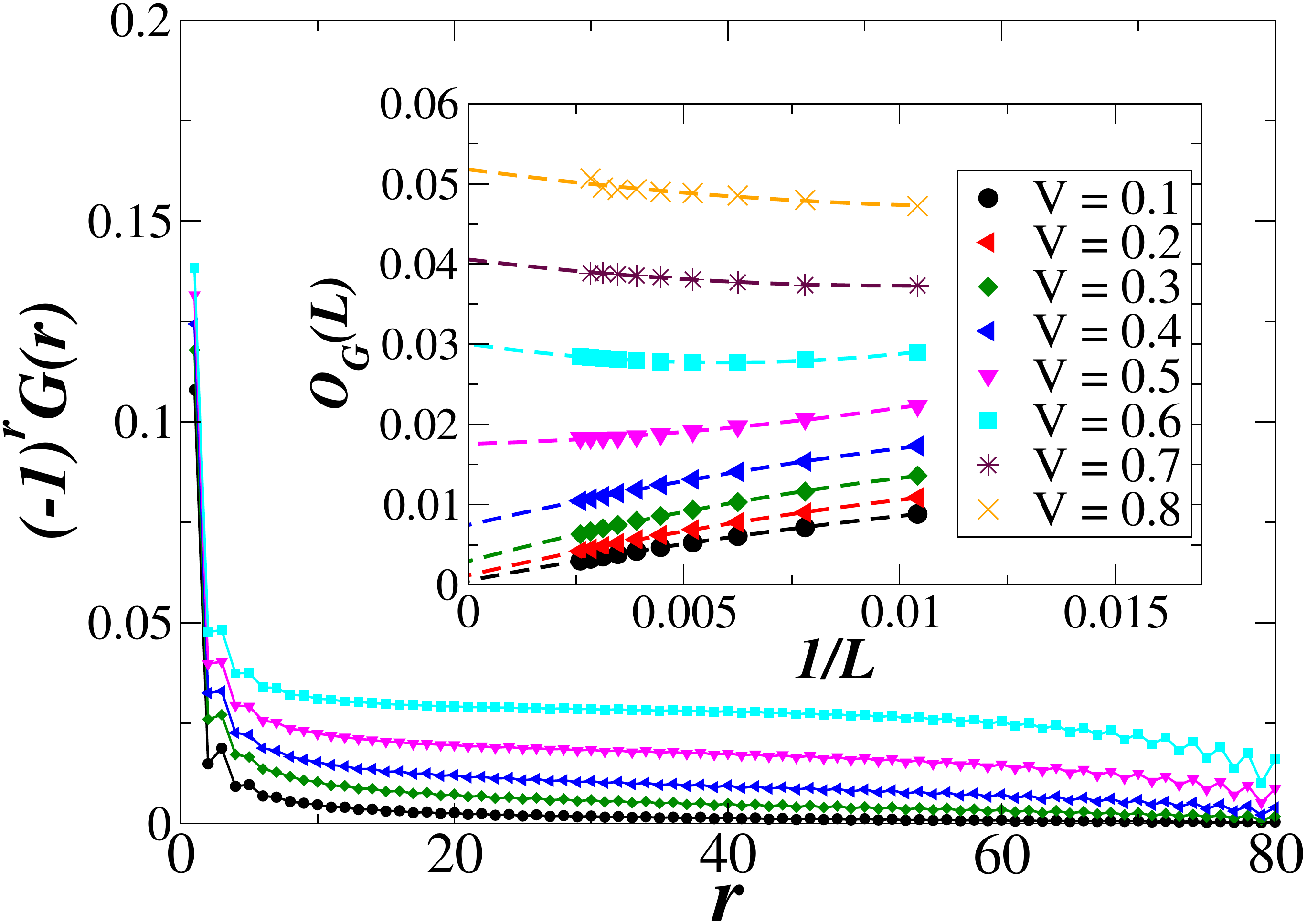}}
\caption{ Plot of correlation function $(-1)^rG_1(r)$, as a function of $r$, at $U = 2.0$ and different values of $V$:
$V=0.1$ (circle), $V=0.3$ (diamond), $V=0.4$ (triangle left),  $V=0.5$ (triangle down) and $V=0.6$ (square).
Inset shows, finite size scaling of  $O_G$, at $U=2.0$ and different values of $V$.}
\label{fig6}
\end{figure}
%*********************************************************************  
 We find pair correlation function, $P(r)$, decay algebraically up to $V=0.4\pm0.05$ for $U=2.0$.
For $V\ge0.4\pm0.05$, the pair correlation function decays exponentially,
 indicating transition from PSF to DW phase.

In Fig.8, we have plotted correlation function, $G_{1}(r)$, as a function of $r$, 
at $U=2.0$ and for different values of $V$. 
 This shows how the density wave order develops in the 
 chain with increase in repulsive interaction, $V$,
 while going from PSF to DW phase. 
 In the inset of Fig.8, we show extrapolation of, $O_G(L)$,
 as a function of 1/L for different values of $V$, and for $U= 2.0$.
From extrapolation of $O_G(L)$, it seems that for $V\gtrsim 0.3$, $O_G$ 
takes finite value for $U=2$. As discussed in the SF to DW transition,
from correlation function, $G_1(r)$, and finite size scaling of  $O_G$,
 we find  that density wave order exists in each of the  chains for $V=0.4\pm0.08$.
  As shown in Fig.8 and the inset of Fig.4, density wave order  develops
 in each of the  chains faster and stabilizes at much lower values of $V$, for $U=2.0$ (Fig.8)
 compared to $U=0.5$ (Fig.4). We find continuous transition from PSF phase to DW phase,
 we did not find PSS phase within our  error-bar.

\subsection{Dimerization}

With  increase in attractive interaction, $U$, between the chains,
 bosons  makes  bound pairs along the rung, while, due to 
 repulsive interaction  $V$, these bound pairs try to avoid each other. 
As a result, in the large limit of $U$ and $V$, positions of the hard core
 bosons in each of the  chains become  strongly correlated. 
  In this limit, the density wave of  each of the chains are correlated to each other.
 To find this correlation in density waves of chains, we have calculated dimer-dimer
 correlation $D(r)=\ \left \langle S^z_{1,0}S^z_{2,0} S^z_{1,r}S^z_{2,r}\right\rangle$, 
where $1$ and $2$ stand for chain indices of the  ladder and $r$  is the distance
 from the middle site of the ladder. As shown in Fig.9, we have plotted 
 $D(r)$ with distance $r$ for $V = 2.2$, and different values of $U$.
For $U=0$, the two chains behave independently, and with increase in
  $U$, we find that the  correlation in density wave increases. As already mentioned,
 increase in $U$ forces bosons to make
 bound pairs along the rungs. The number of boson pairs in terms of spins,
 can be defined as, N$_{pair} \ = \ \sum_i<s^z_{1,i}s^z_{2,i}> /\frac{L}{4}$,
 where $i$ is the site index of the chain.
 For  small values of $U$, since the system has large fluctuation effects,
 the number of  pairs is quite small. In fact, in this limit, the system has
  loosely bound pairs along the rungs. While,  with increase in 
 $U$, $N_{av}$ increases, displaying crossover of the system to strongly bound pairs.
 We  also find that, repulsive interaction, $V$,  helps to stabilize these bound pairs.
 This is shown in the inset of Fig.9, where we have plotted $N_{av}$ verses 
 $U$, for different values of $V$.
%********************************Fig.9.*****************************
\begin{figure}[h]
\rotatebox{0}{\includegraphics*[width=\linewidth]{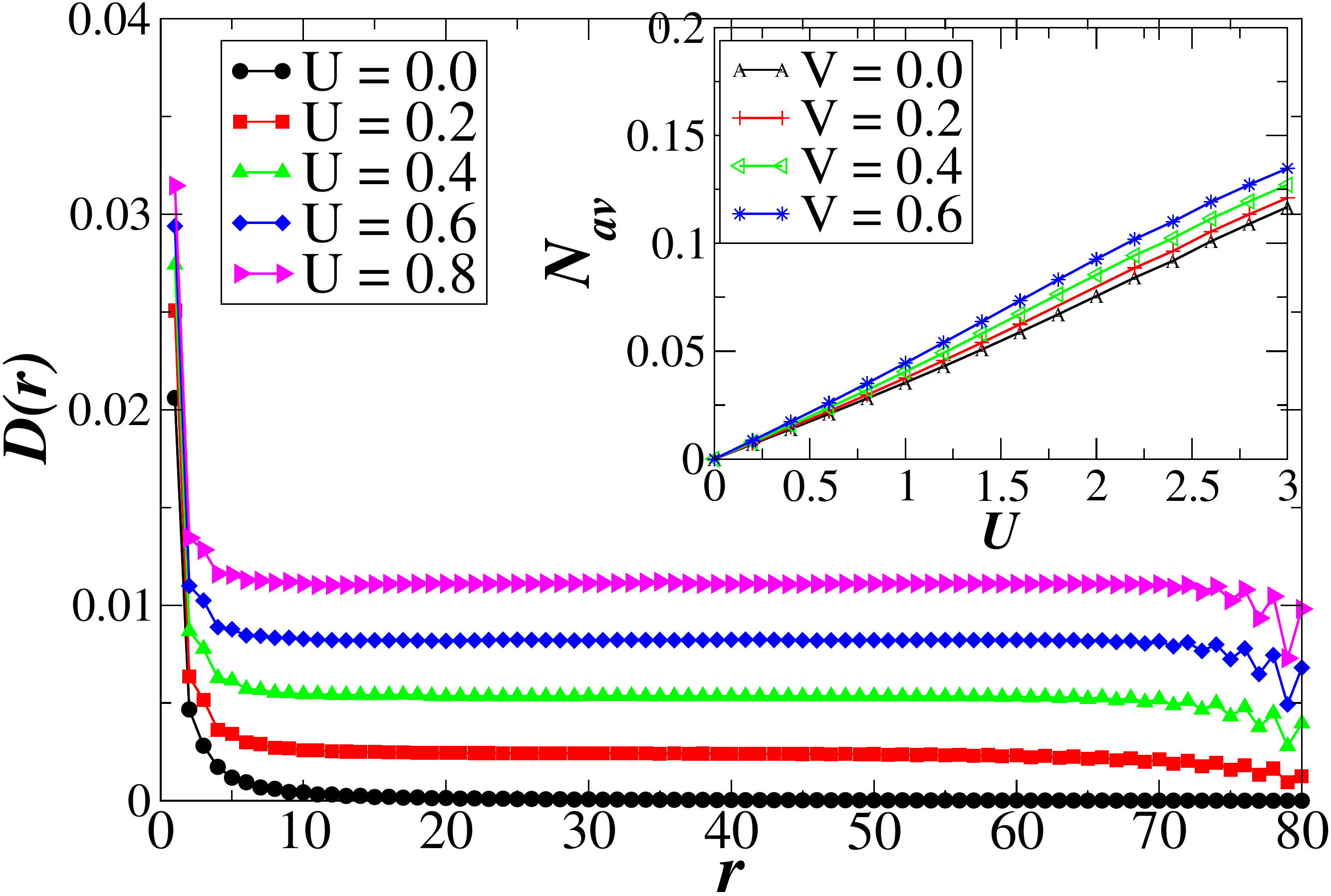}}
\caption{Plot of dimer-dimer correlation function $D(r)$, as function of $r$, for $V=2.2$ and different values of $U$. 
Inset shows plot of $N_{av}$ as a function of $U$, and  different values of $V$.}
\label{fig9}
\end{figure}

%*************************************************************

As we have discussed, in the large $U$ and $V$ limit, the  system forms a 
density wave of strongly bound pairs and positions of hard-core bosons in each
 of the chains become strongly correlated\cite{as,safi}.  In spin language, spins align ferromagnetically
 along the rung of the ladder, while, antiferromagntically along  each of the  chains,
 (as shown in the schematic of Fig.10(a)).
To show this, in Fig.10(a), we have plotted spin-density, $\langle S_l^z\rangle$,
 of the ladder with position $l$, for $U=2$, and $V=1.5$. For a clear view of $\langle S_l^z\rangle$,
 numbering of $l$ index is done in a different way, which is shown in schematic of Fig.10(a).
  Spin-density, $<S_l^z>$, along the rungs takes 
same value and are in the same direction, while, along  the chains,
they are oriented in opposite directions. Such a configuration with parallel
 spin within each rungs and anti-parallel spin along each chain of the ladder
 structure can be represented as $|\uparrow \uparrow \downarrow \downarrow \uparrow \uparrow...\rangle$. 
In hardcore bosonic language, due to attractive interaction, $U$, 
hardcore bosons form bound pairs along the rungs, while, due to repulsive interaction, $V$,
present in each of the chains, these rung pairs try to avoid each other. As a result, these rung pairs
reside on alternate rungs and this configuration can be represented by
   $|110011..\rangle$. As shown in Fig.10(b), this configuration 
can also be visualised by looking at the density correlation function, $g(l) = \langle s^z_0s^z_l\rangle$,
for both the chains (full ladder),
 where $s^z_0$ is considered as the  middle spin site of the ladder.
Numbering of $l$ index for $g(l)$, as shown schematically in Fig.10(a), is done differently compared to $G_1(r)$.
As shown in Fig.10(b), periodicity of the density wave on the ladder is twice the lattice spacing. 
 The structure factor, defined as, $G(k)= \frac{1}{L/2}\sum_l\exp(ik\cdot l)<s^z_os^z_l>$, has  peaks at 
$-\pi/2$ and $\pi/2$.
%********************************Fig.10.*****************************
\begin{figure}[h]
\rotatebox{0}{\includegraphics*[width=\linewidth]{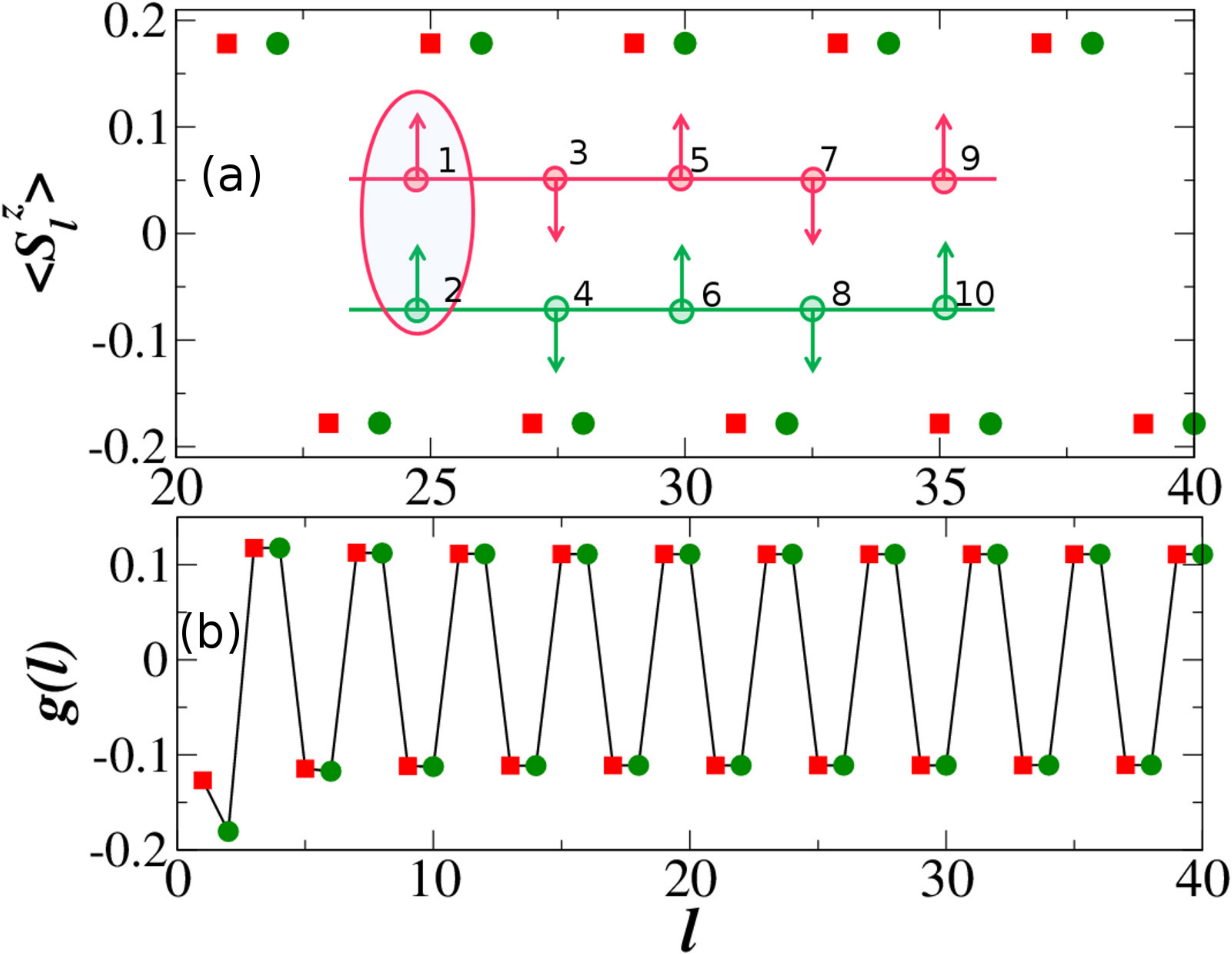}}
\caption{(a) Plot of spin-density $<s^z_l>$, as function of $l$, for $U=2.0$ and $V=1.5$.
 Spin-density of the first chain is denoted with a square, while the second chain is denoted with a circle.
 Inset shows the schematic of dimerization of spins in two chains.
(b) Plot of density correlation $g(l)$, as a function of $l$, for $U=2.0$ and $V=1.5$.}   
\label{fig5}
\end{figure}

%*************************************************************
\section{Conclusion}
In summary, we have studied various phases of hardcore bosons in two coupled chains,
 with inter chain attraction and inter chain nearest neighbor repulsion between the bosons.
We find that the ground state phase diagram has mainly three phases, SF, PSF and DW.
We have estimated the phases and the phase boundaries accurately through appropriate two 
body and four body correlation functions and at times the corresponding structure factors.  
The model discussed in this article is a simplified description of bilayer dipolar bosons with
 dipole moments perpendicular to the plane. 
 Although, we truncated the long range dipolar interaction to nearest neighbour repulsion, 
this model contains essential ingredients for the formation of `pair superfluid' and `pair density wave' phases.
Similar to the BCS-BEC crossover of fermions, in this system, bosons can 
undergo a transition from a weakly bound paired superfluid state 
to density wave of strongly bound pairs.
\section{Acknowledgments}
We would like to thank A. V. Mallik, Diptiman Sen and Subroto Mukerjee for fruitful discussions.
B.P. thanks the UGC, Govt. of India for support through fellowship and S.K.P. acknowledges DST, Govt. of India for financial support.


\begin{thebibliography}{99}
\bibitem{Thierry}T. Lahaye, T. Koch, B. Fr\"ohlich, M. Fattori, J. Metz, A. Griesmaier,
 S. Giovanazz, T. Pfau, {Nature {\bf{448}}, 672 (2007).}
\bibitem{Lu}M. Lu, N. Q. Burdick, S. H. Youn, and B. L. Lev, {Phys. Rev. Lett. {\bf{107}}, 190401 (2011).}
\bibitem{weim} R. L\"ow, H. Weimer, J. Nipper, J. B Balewski, B. Butscher, H. P. B\"uchler, and T. Pfau, {J. Phys. B. {\bf{45}},113001 (2012).}
\bibitem{BoYan} B. Yan, S. A. Moses, B. Gadway, J. P. Covey, K. R. A. Hazzard, A. M. Rey, D. S. Jin, and J. Ye, {Nature {\bf{501}}, 521 (2013).}
\bibitem{Markus} M. Greiner, C. A. Regal, and D. S. Jin, {Nature {\bf{426}}, 537 (2003).}
\bibitem{regal} C. A. Regal, M. Greiner, and  D. S. Jin, {Phys. Rev. Lett. {\bf{92}}, 040403 (2004).} 
\bibitem{Ni} K.-K. Ni, S. Ospelkaus, M. H. G. de Miranda, A. P\'eer, 
B. Neyenhuis, J. J. Zirbe, S. Kotochigova, P. S. Julienne, D. S. Jin, and J. Ye,
{Science {\bf{322}}, 231 (2008).}
\bibitem{kuk} A. Kuklov, N.  Prokof'ev, and B. Svistunov, {Phys. Rev. Lett. {\bf{92}}, 050402 (2004).}
\bibitem{chi}C. Chung, S. Fang, and P. Chen, {Phys. Rev. B. {\bf{85}}, 214513 (2012).}
\bibitem{Tre}C. Trefzger, C. Menotti, and M. Lewenstein, {Phys. Rev. Lett. {\bf{103}},
 035304 (2009).} 
\bibitem{safi}A. Safavi-Naini, S. G. S\"oyler, G. Pupillo, H. R. Sadeghpour, and B. Capogrosso-Sansone, {New J. Phys. {\bf{15}}, 013036 (2013).}
\bibitem{Amacia}A. Macia, G. E. Astrakharchik, F. Mazzanti, S. Giorgini, and J. Boronat,
 {Phys. Rev. A. {\bf{90}}, 043623 (2014).}
\bibitem{cat} J. Catani, L. De Sarlo, G. Barontini, F. Minardi, and M. Inguscio,
 {Phys. Rev. A. {\bf{77}}, 011603(R) (2008).}
\bibitem{Hong}H. C. Jiang, L. Fu, and C. K. Xu, {Phys. Rev. B. {\bf{86}}, 045129 (2012)} 
\bibitem{kim}E. Kim, and M. H. W. Chan, {Nature {\bf{427}}, 225 (2004)}
\bibitem{Duk} D. Y. Kim, and M. H. W. Chan, Phys. Rev. Lett. {\bf{109}}, 155301 (2012)
\bibitem{dlkov}  D. L. Kovrizhin, G. V. Pai, and S. Sinha, {Eur. phys. Lett. {\bf{72}}, 162 (2005) }
\bibitem{Melko}R. G. Melko, A. Paramekanti, A. A. Burkov, A. Vishwanath, D. N. Sheng, and L. Balents, {Phys. Rev. Lett. {\bf{95}}, 127207 (2005).}
\bibitem{Sengupta}P. Sengupta, L. P. Pryadko, F. Alet, M. Troyer, and G. Schmid,
 {Phys. Rev. Lett. {\bf{94}}, 207202 (2005).}
\bibitem{sinha}S. Sinha, and L. Santos, {Phys. Rev. Lett. {\bf{99}}, 140406 (2007).} 
\bibitem{jiang}H. C. Jiang, M. Q. Weng, Z. Y. Weng, D. N. Sheng, and L. Balents, 
{Phys. Rev. B. {\bf{79}}, 020409(R) (2009).}
\bibitem{adam}A. B\"uhler, and H. P. B\"uchler, {Phys. Rev. A. {\bf{84}}, 023607 (2011).}
\bibitem{N.Henke}N. Henkel, R. Nath, and T. Pohl, {Phys. Rev. Lett. {\bf{104}}, 195302 (2010).}
\bibitem{F.Cinti}F. Cinti, P. Jain, M. Boninsegni, 
A. Micheli, P. Zoller, and G. Pupillo, {Phys. Rev. Lett. {\bf{105}}, 135301 (2010).}
\bibitem{Xi}X. Li, W. V. Liu, and C. Lin, {Phys. Rev. A. {\bf{83}}, 021602(R) (2011).}
\bibitem{w.zhang} W. Zhang, R. Yin, and Y. Wang, {Phys. Rev. B. {\bf{88}}, 174515 (2013).} 
\bibitem{giamarchi}T. Giamarchi, Quantum physics in One dimension (Clarendon  Press, Oxford UK, 2004).
\bibitem{T.D}T. D. k\"uhner and H. Monien {Phys. Rev. B. {\bf{58}}, R14741(R) (1998).}
\bibitem{Till}T. D. k\"uhner, S. R. White and H. Monien, {Phys. Rev. B. {\bf{61}}, 12474 (2000).}
\bibitem{G.Roux}G. Roux, T. Barthel, I. P. McCulloch, C. Kollath, U. Schollw\"ock, and T. Giamarchi,
 {Phys. Rev. A. {\bf{78}}, 023628 (2008).}
\bibitem{M.A} M. A. Cazalilla, R. Citro, E. Orignac, and M. Rigol, 
{Rev. Mod. Phys. {\bf{83}}, 1405 (2011).}
\bibitem{S.E}S. Ejima, F. Lange, H. Fehske, F. Gebhard,
 and K. zu M\"unster, {Phys. Rev. A. {\bf{88}}, 063625, (2013).}
\bibitem{Meetu}M. S. Luthra, T. Mishra, R. V. Pai, and B. P. Das,
 {Phys. Rev. B. {\bf{78}}, 165104 (2008).}
\bibitem{arg}A. Arg\"uelles, and L. Santos, {Phys. Rev. A. {\bf{75}}, 053613 (2007).}
\bibitem{Anzi}A. Hu, L. Mathey, I. Danshita, E. Tiesinga, C. J. Williams, and C. W. Clark,
 {Phys. Rev. A. {\bf{80}}, 023619 (2009).}
\bibitem{lmat}L. Mathey, I. Danshita, and C. W. Clark, {Phys. Rev. A. {\bf{79}}, 011602(R) (2009).}
\bibitem{lmazza}L. Mazza, M. Rizzi, M. Lewenstein, and J. I. Cirac, 
{Phys. Rev. A. {\bf{82}}, 043629 (2010).}
\bibitem{bat}G. G. Batrouni, R. T. Scalettar, V. G. Rousseau, and B. Gr\'emaud, 
{Phys. Rev. Lett. {\bf{110}}, 265303 (2013).} 
\bibitem{swap}A. Kundu and S. K. Pati, {Eur. Phys. Lett. {\bf{85}}, 43001 (2009).}
\bibitem{tapan1}T. Mishra, R. V. Pai, S. Mukherjee, A. Paramekanti, 
{Phys. Rev. B. {\bf{87}}, 174504 (2013).}
\bibitem{tapan}T. Mishra, R. V. Pai, S. Mukherjee, {Phys. Rev. A. {\bf{89}}, 013615 (2014).}
\bibitem{mish}T. Mishra, J. Carrasquilla and M. Rigol, {Phys. Rev. B. {\bf{84}}, 115135 (2011).}
\bibitem{gs} G. S\"oyler, B. Capogrosso-Sansone, N. V. Prokof'ev, B. V. Svistunov,
 {New J. Phys.{\bf{11}}, 073036 (2009).}
\bibitem{mg} M. Guglielmino, V. Penna, and B. Capogrosso-Sansone, {Phys. Rev. A. {\bf{84}}, 031603(R)(2011).}
\bibitem{as} A. Safavi-Naini, B. Capogrosso-Sansone, and A. Kuklov, {Phys. Rev. A. {\bf{90}}, 043604 (2014).}
\bibitem{man} M. Singh, T. Mishra, R. V. Pai and B. P. Das, {Phys. Rev. A. {\bf{90}}, 013625(2014).}
\bibitem{White}S. R. White, {Phys. Rev. Lett. {\bf{69}}, 2863 (1992)};
{Phys. Rev. B {\bf{48}}, 10345 (1993).}  
\bibitem{Schollwock} U. Schollw\"ock, {Rev. Mod. Phys. {\bf{77}}, 259 (2005).}
\bibitem{Rpai} R. V. Pai, R. Pandit, H. R. Krishnamurthy, and S. Ramasesha,
 {Phys. Rev. Lett. {\bf{76}}, 2937 (1996).}
\bibitem{Rpai1}R. V. Pai and R. Pandit, {Phys. Rev. B. {\bf{71}}, 104508 (2005).}
\bibitem{G.GBAT1}G. G. Batrouni, R. T. Scalettar, G. T. Zimanyi, and A. P. Kampf,
 {Phys. Rev. Lett. {\bf{74}}, 2527 (1995).}
\bibitem{D.ross} D. Rossini and R. Fazio, {New J. Phys. {\bf{14}}, 065012 (2012).}
\bibitem{J.M.K}J. M. Kosterlitz and D. J. Thouless, {J. Phys. C {\bf{6}}, 1181 (1973).}
\end{thebibliography}
\end{document}